\documentclass[
aps
,prb
,twocolumn,10pt
,superscriptaddress
,longbibliography
,amsmath
,amssymb
,amsfonts
,floatfix
]{revtex4-2}

\usepackage{float}
\usepackage{commath}
\usepackage{graphicx}
\usepackage{verbatim}

\usepackage{color}
\usepackage[dvipsnames,svgnames,x11names,hyperref]{xcolor}

\usepackage[normalem]{ulem}

\usepackage[normalem]{ulem}
\usepackage{times}

\usepackage[colorlinks=true,
            linkcolor=linkcolor,
            urlcolor=linkcolor,
            citecolor=linkcolor,
            unicode,
            pdfencoding=auto]{hyperref}
            
\usepackage{siunitx}

\definecolor{linkcolor}{RGB}{6,69,173} 
\definecolor{diffcolor}{RGB}{175,31,36} 

\def\beq {\begin{equation}}
\def\eeq {\end{equation}}
\def\beqar {\begin{eqnarray}}
\def\eeqar {\end{eqnarray}}

\newcommand{\soleil}{Synchrotron SOLEIL, L'Orme des Merisiers, Saint-Aubin, BP 48, F-91192 Gif-sur-Yvette, France}

\newcommand{\lps}{Universit\'e Paris-Saclay, CNRS, Laboratoire de Physique des Solides, 91405, Orsay, France}

\begin{document}

\title{Resonant X-ray spectroscopies on Chromium 3\textit{d} orbitals in CrSBr}

\author{Victor Por\'{e}e}
\email{victor.poree@synchrotron-soleil.fr} 
\affiliation{\soleil}

\author{Alberto Zobelli}
\affiliation{\soleil}
\affiliation{\lps}

\author{Amit Pawbake}
\affiliation{Laboratoire National des Champs Magn\'{e}tiques Intenses, CNRS-UGA-UPS-INSA-EMFL, 38042, Grenoble, France}

\author{Jakub Regner}
\affiliation{Department of Inorganic Chemistry, University of Chemistry and Technology Prague, Technicka 5, 166 28 Prague 6, Czech Republic}

\author{Zdenek Sofer}
\affiliation{Department of Inorganic Chemistry, University of Chemistry and Technology Prague, Technicka 5, 166 28 Prague 6, Czech Republic}

\author{Cl\'{e}ment Faugeras}
\affiliation{Laboratoire National des Champs Magn\'{e}tiques Intenses, CNRS-UGA-UPS-INSA-EMFL, 38042, Grenoble, France}

\author{Alessandro Nicolaou} 
\email{alessandro.nicolaou@synchrotron-soleil.fr} 
\affiliation{\soleil}


\begin{abstract}
We investigate the Cr electronic structure and excitations in CrSBr, a layered magnetic semiconductor, using a combination of resonant x-ray spectroscopic techniques. X-ray absorption spectroscopy (XAS) and resonant inelastic x-ray scattering (RIXS) spectra collected at the Cr $L_{2,3}$ edges reveal significant linear dichroism, which arises from the distorted octahedral environment surrounding the Cr$^{3+}$ ions. The origin of the bright excitons observed in this compound is examined through a comparison of the d-d excitations identified in the RIXS spectra, the x-ray excited optical luminescence (XEOL) spectra, and previously reported optical spectroscopic and theoretical studies. To further understand these phenomena, we develop a multiplet model based on a crystal electric field (CEF) approach that accounts for the local environment of Cr ions. This model successfully reproduces several experimental features, while also suggesting strong hybridization effects between Cr 3\textit{d} orbitals and ligands that are not fully captured by the present framework. These findings advance our understanding of the electronic structure and excitonic behavior in CrSBr and provide a foundation for future \textit{in-situ} and \textit{operando} studies of CrSBr-based devices for spintronic and optoelectronic applications.
\end{abstract}


\maketitle


\section{Introduction}
In the ever-evolving landscape of materials science, the discovery and exploration of novel phases hold immense promise for technological advancements. Among the various kinds of materials with potential applications, van der Waals (vdW) compounds are extremely promising due to their high tunability. The ability to control the number of atomically thin layers and the possibility of constructing twisted homolayers and/or heterostructures allow tailoring the characteristics of many of these systems~\cite{TMLMAG,MoS2_vdW_multilayers,WSe2_regan,Hetero_vdW_science}. Particularly, the emerging field of vdW magnets offers new opportunities to advance both our understanding of 2D magnetism and the creation of new devices~\cite{FGT_monolayer,CRXGe_magnonInsul,MnSe2_ferro,VSe2_ferro}. Among these remarkable materials, Cr-based vdW magnets have emerged as particularly promising, exhibiting properties such as magneto-optical effects and giant tunneling magnetoresistance~\cite{Nat_Phy_CrI3,CrI3_PL_Peng,nano_lett_CrBr3_PL,CrPS4,CrSBr_magnetoRes,CrSBr_Nature2023}. Materials with tunable photoluminescence via control of their magnetic states are of great interest to researchers in the fields of magnetism, optoelectronics, and spintronics.

CrSBr is one such material, consisting of monolayers stacked along the crystallographic \textbf{c} axis and held together via the van der Waals interaction. Each monolayer comprises square lattices formed by edge-sharing [CrS$_4$Br$_2$] distorted octahedra~\cite{CrSBr_magstruct_PSI}. Two square sub-lattices can be distinguished, one shifted with respect to the other by half a unit cell along both the \textbf{a} and \textbf{b} directions and flipped upside down. This arrangement connects the square sub-lattices through their S ions, with the monolayers facing each other with opposing Br ions, obeying an overall orthorhombic $P_{mmn}$ space group symmetry. As described before, the Cr$^{3+}$ cations are surrounded by four S$^{2-}$ and two Br$^{1-}$ anions forming a distorted octahedron. Consequently, the Cr ions experience a C$_{2v}$ crystal field environment, which should lift the degeneracy of the five $d$ orbitals~\cite{Wang_CrSBr,CrSBr_NatComm_Klein}, with a predicted strong hybridization with its ligands. Due to the higher energy of a doubly occupied orbital relative to the crystal field splitting, Cr$^{3+}$ exhibits a ground state with three singly occupied \textit{d} orbitals corresponding to a high spin S~=~3/2 configuration~\cite{GOSER1990129,CrSBr_magstruct_PSI}. The magnetic moments seem to develop in-plane ferromagnetic correlations starting from 160~K~\cite{CrSBr_magstruct_PSI,CrSBr_Lee_2021}. However, no net magnetic moment is observed in bulk samples down to approximately 140~K, nor any significant anomaly in magnetic susceptibility measurements, raising questions about the establishment of intra-layer short-range or long-range order ~\cite{Triple_stage_mag_trans,CrSBr_Long}. Below approximately 140~K, the strong intra-layer ferromagnetic interaction and weak antiferromagnetic coupling between the layers stabilize a long-range A-type antiferromagnetic structure, with the magnetic moments pointing along the \textbf{b} crystallographic axis ~\cite{GOSER1990129,CrSBr_magstruct_PSI,CrSBr_Lee_2021}. Magnetization and neutron diffraction studies indicate that the \textbf{b} axis is the magnetic easy axis, while the \textbf{a} and \textbf{c} axes are intermediate and hard axes, respectively~\cite{CrSBr_magnetoRes,CrSBr_magstruct_PSI,CrSBr_Lee_2021}. This behavior reflects the ferromagnetic coupling between Cr nearest neighbours and a strong planar anisotropy~\cite{CrSBr_Scheie_magnons,CrSBr_magstruct_PSI,CrSBr_Lee_2021,CrSBr_magnetoRes,Wang_CrSBr,CrSBr_NatComm_Klein}. The in-plane magnetic easy axis observed in CrSBr contrasts with some other Cr-based vdW magnets, such as the trihalides where the magnetic moments tend to be oriented along the stacking direction~\cite{Nat_Phy_CrI3,nano_lett_CrBr3_PL,CrPS4_MagStruct}. However, some aspects of the magnetic structure remain elusive, such as the possible existence of topological edge modes related to a proposed Dzyaloshinskii-Moriya interaction~\cite{CrSBr_Scheie_magnons}, as well as some contradicting observations of magnetic transitions around 40~K~\cite{CrSBr_magstruct_PSI,CrSBr_Long}. Additionally, its air stability, optical band gap of 1.8~eV combined with a large magnetoresistance make CrSBr a promising material for applications~\cite{NanoLetter_Ziebel}. 

Of topical interest is the coupling of optical transitions with the underlying magnetic structure. For instance, studies on CrI$_3$ and CrBr$_3$ have shown that photoluminescence dichroism can be reversed by flipping the magnetization via the application of a magnetic field~\cite{Nat_Phy_CrI3,CrI3_PL_Peng,nano_lett_CrBr3_PL}. A similar procedure can be applied to CrSBr, where its photoluminescence experiences a redshift while the system transitions from an antiferromagnet to a ferromagnet~\cite{CrSBr_NatMat,CrSBr_NatNano}. This effect is further tunable under pressure~\cite{pawbake2023magnetooptical}, also revealing a spin-flop transition. In CrSBr, the photoluminescence is attributed to the lowest bright Wannier excitonic transition~\cite{Wang_CrSBr}. The observed redshift could be captured via GW-BSE calculation~\cite{CrSBr_NatMat}, indicating that within the AFM state, the exciton remains confined to the vdW monolayers, extending widely in the \textbf{ab} plane but narrowly along the \textbf{c}-axis due to a spin-forbidden nearest neighbor inter-layer hopping. Interestingly, this constraint relaxes when the system transitions to the FM state, as the inter-layer hybridization becomes spin-allowed. As a consequence of this hybridization, a band splitting occurs at both sides of the band gap, reducing its width and consequently lowering the exciton’s energy. Additionally, the exciton displays a strong optical anisotropy, being visible for incident polarization along the \textbf{b}-axis and absent in the \textbf{a}-axis, further suggesting strong constraints on the allowed and forbidden electronic transitions~\cite{CrSBr_NatMat}. As supported by theory, the electronic states with significant Cr 3\textit{d} character play a major role in the observed magneto-optic effect, motivating the need for further spectroscopic characterisation~\cite{Linhart}.

To gain a deeper understanding of the Cr$^{3+}$ electronic states in CrSBr, particularly the crystal electric field (CEF) splitting of its 3\textit{d} orbitals, we conducted x-ray absorption spectroscopy (XAS), resonant inelastic x-ray scattering (RIXS), and x-ray excited optical luminescence (XEOL) measurements at the Cr $L_{2,3}$ edges. The resonant nature of these techniques enhances sensitivity to Cr-dominated states, allowing us to develop a CEF model based on multiplet calculations. This study paves the way for future \textit{in-situ} and \textit{operando} investigations of CrSBr-based devices using advanced spectroscopic methods.

\section{Methods}
\begin{figure}
\includegraphics[width=0.48\textwidth]{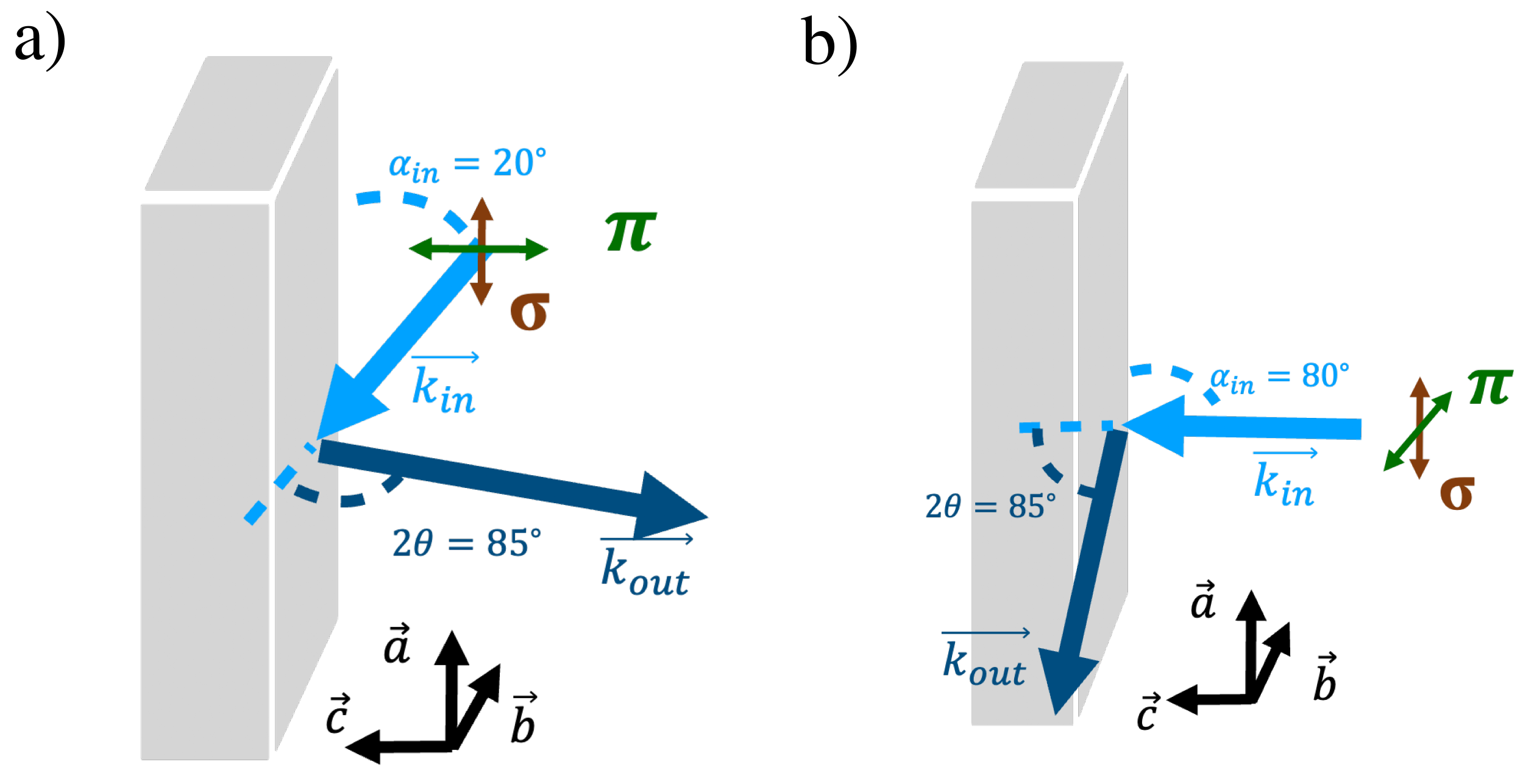}
\centering
\caption{Experimental geometries in grazing (a) and normal (b) incidence. The light and dark blue arrows represent the incident and outgoing beams, respectively. Crystallographic axis are given with the \textbf{a} and \textbf{b} axes within the sample's flat surface.}
\label{Exp_geo}
\end{figure}

High-quality CrSBr crystals were synthesized by chemical vapor transport~\cite{pawbake2023magnetooptical}, producing flat samples with the \textbf{a} and \textbf{b} crystallographic axes lying within the plane, with the crystals noticeably longer along the \textbf{a}-axis. Raman spectroscopy confirmed the quality of the samples (see Sup. Mat. and \cite{Amit_PRB}). A large, plate-like sample was selected and glued to a copper gold-plated Omicron-type sample holder using silver paint. The sample was cleaved in air using scotch tape immediately before being introduced in the load lock and subsequently transferred into the experimental chamber. The experiment was performed at the inelastic branch of the SEXTANTS beamline \cite{SACCHI2013} of SOLEIL employing the AERHA spectrometer \cite{CHIUZBUAIAN2014} and the MAGELEC sample environment \cite{MAGELEC}. An experimental geometry with the crystallographic \textbf{a}-axis vertical and the \textbf{b} and \textbf{c} axes within the horizontal scattering plane was selected (illustrated in Fig.~\ref{Exp_geo}\textbf{a}). Measurements using linearly polarized incident radiation (horizontal, $\pi$, and vertical, $\sigma$, orientations of the incident radiation's electric field) were conducted at both grazing and \textit{quasi}-normal incidence (beam arriving at approximately 20$^\circ$ and 80$^\circ$ with respect to the surface plane, respectively, see Fig.~\ref{Exp_geo}\textbf{ab}). A RIXS scattering angle of $2\theta$~=~85$^\circ$ was used to minimize elastic contributions for $\pi$ incident polarization. Multiple XAS (via total electron yield) and RIXS spectra were collected with incident synchrotron radiation covering both the $L_{3}$ and $L_{2}$ edges of Cr. The overall energy resolution of the RIXS spectra was estimated at approximately 220~meV. The MAGELEC sample environment enabled both XAS and RIXS measurements at temperatures ranging from 300~K down to 20~K. XEOL spectra were acquired using an in-house optical system coupled through an optical fiber to a state an optical spectrometer (Princeton Isoplane Advanced) available at the RIXS end-station of the SEXTANTS beamline. The photoluminescence was triggered by irradiation of the sample with synchrotron radiation, with incident energies close the $L_{3}$ edge of Cr. XEOL spectra were recorded at 20~K for various incident polarizations, with a resolution of approximately 20~meV. Finally, the recorded XAS and RIXS spectra were modeled via multiplet cluster calculations using the quantum many body script language named Quanty~\cite{Quanty1,Quanty2,Quanty3}.

\section{Results and Discussion}
\subsection{X-ray Absorption spectroscopy}\label{XAS_section}
\begin{figure}
\includegraphics[width=0.48\textwidth]{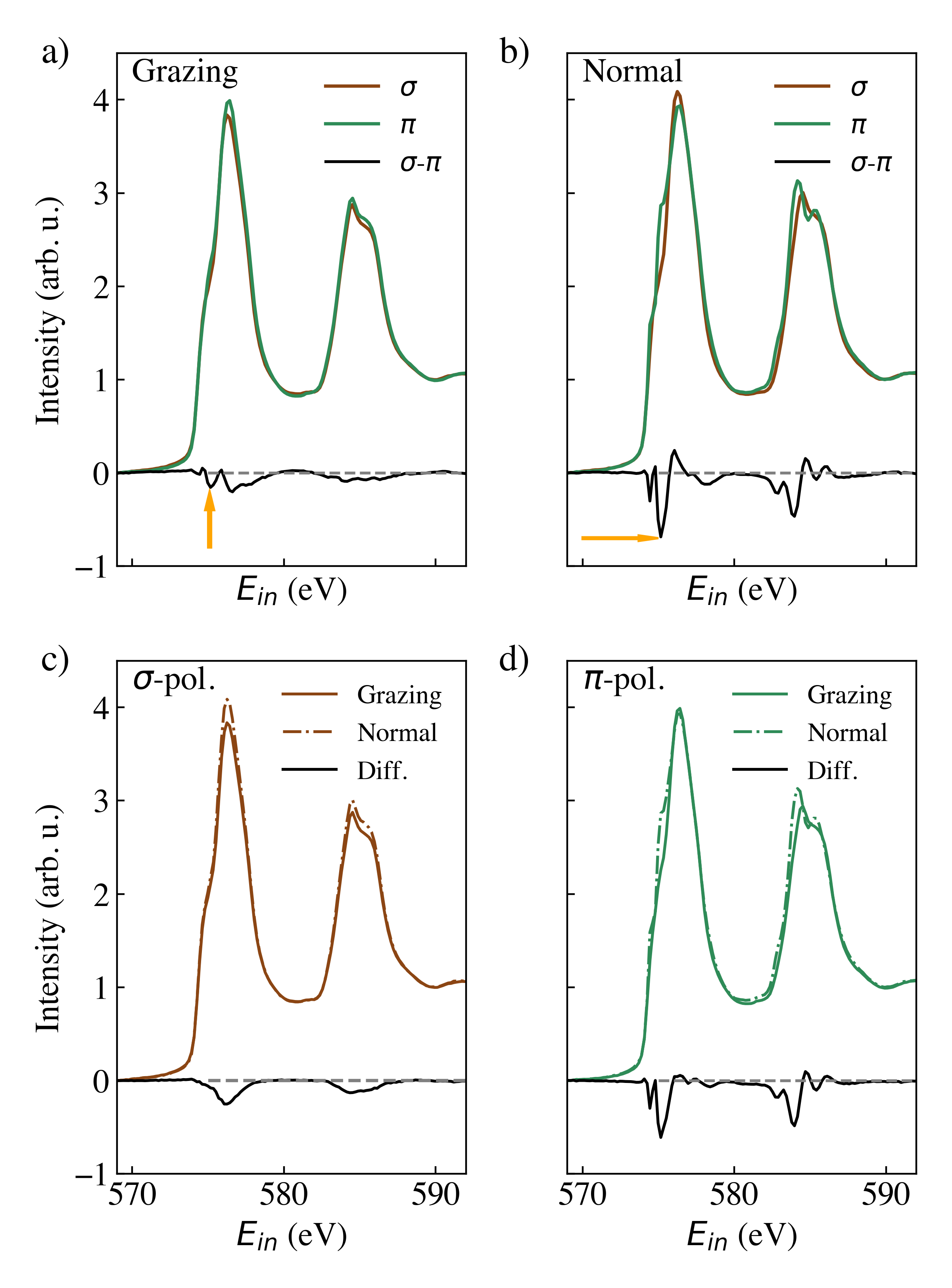}
\centering
\caption{XAS spectra and associated linear dichroism collected at RT for both grazing (a) and normal (b) incidence at the Cr $L_{2,3}$ edges. The $\pi$ ($\sigma$) polarization corresponds to an incident radiation's electric field oriented parallel (perpendicular) to the scattering plane. The orange arrows indicate the incident energy used to produce the spectra displayed in Fig.~\ref{Fig_RIXS_LD1}. Comparison between spectra collected in grazing and normal incidence for $\sigma$ (c) and $\pi$ (d) polarizations.}
\label{Fig_geo_and_LD}
\end{figure}

We begin by examining the x-ray linear dichroism (XLD) measured on CrSBr at room temperature (RT) over the chromium $L_{2,3}$ edges. This technique provides information about the anisotropic electronic distribution around the Cr$^{3+}$ ion, by selectively enhancing transitions from filled 2\textit{p} to empty 3\textit{d} orbitals depending on the incident polarization and energy. As shown in Fig.~\ref{Fig_geo_and_LD}\textbf{a}, the XAS spectra collected in grazing incidence exhibit a weak linear dichroism with maxima (negative in the chosen convention) located at both edges, indicating that transitions to 3\textit{d} orbitals having components perpendicular to the (\textbf{a},\textbf{b}) plane are prevalent compared to in-plane ones. When moving to normal incidence, the linear dichroism signal becomes more structured and increases significantly (see Fig.~\ref{Fig_geo_and_LD}\textbf{b}). We note an inversion of the dichroism at the maximum of the $L_3$ edge as well as a net shift of approximately 0.25~eV of the maximum of the $L_2$ in the $\pi$ polarization (Fig.~\ref{Fig_geo_and_LD}\textbf{d}). Given that the measurement is conducted relatively far from the N\'eel temperature, this spectral shift is unlikely to stem from magnetic exchange interaction. Instead, it could be interpreted as a consequence of the crystal field splitting of $e_\mathrm{g}$ orbitals~\cite{NiO_Haverkort}. The presence of substantial linear dichroism is coherent with the significant distortion of the octahedra around the Cr ions and aligns with recently reported XLD measurements~\cite{CrSBr_XMCD}. Due to the large number of possible 2\textit{p}$^5$3\textit{d}$^4$ eigenstates (1260), a direct interpretation of the data in terms of occupied orbitals is challenging and necessitates detailed modeling~\cite{Cr_XAS_DFT,Hunault_Ruby}.

\subsection{Resonant Inelastic X-ray Scattering}

\begin{figure}[b]
\includegraphics[width=0.49\textwidth]{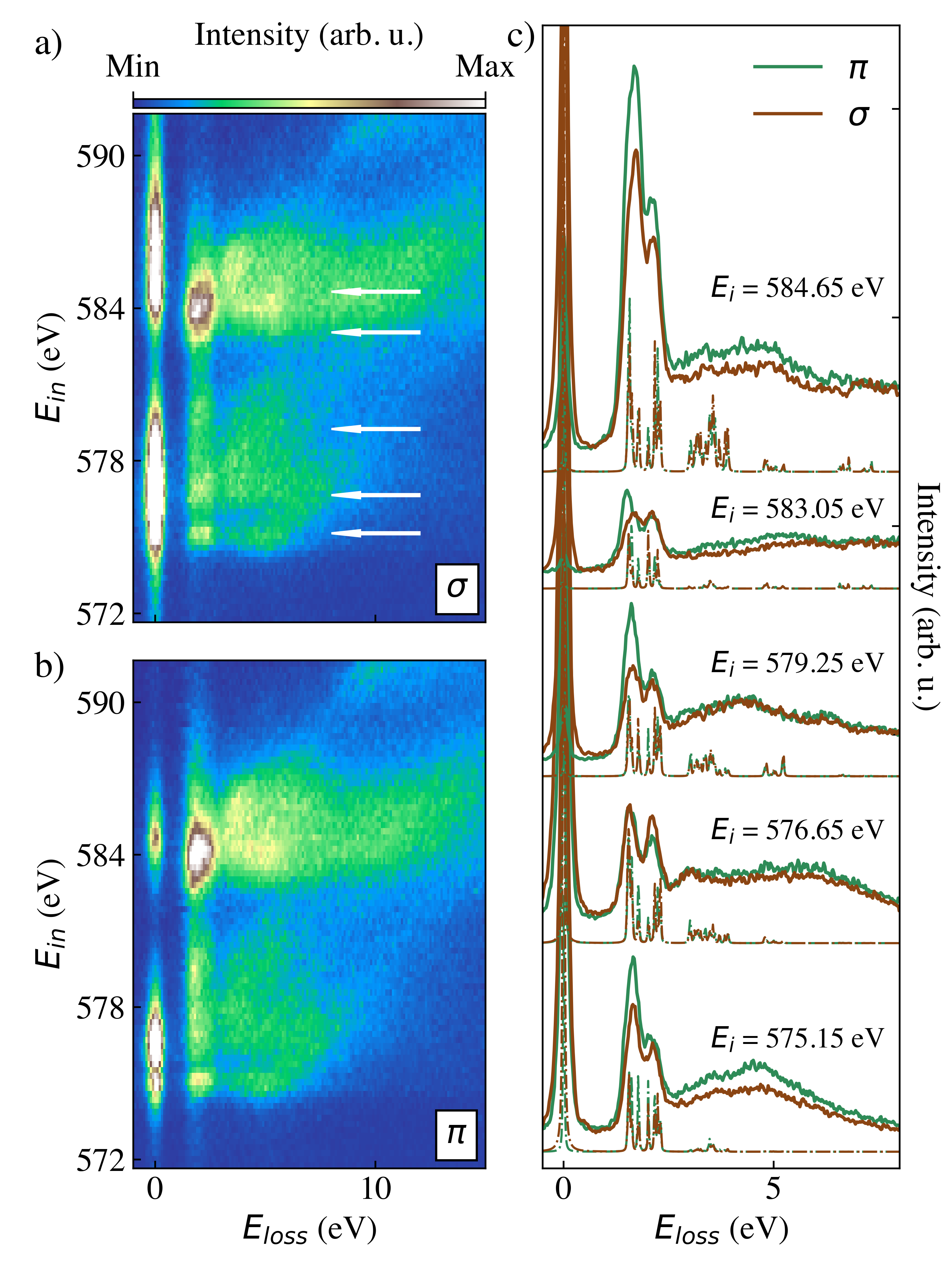}
\centering
\caption{RIXS maps measured at 150~K using linear (a) $\sigma$ and (b) $\pi$ incident polarizations. The black arrows in panel (a) point to the incident energies used for the high statistics spectra of panels (c). (c) RIXS spectra measured in both linear polarization ($\sigma$ in brown and $\pi$ in green) collected at 25~K with incident energies as indicated in panel (a). The curves are shifted upwards with increasing incident energies for better visualization. These measurements were performed at a momentum transfer \textbf{k}~=~(0, 0.152, 0.3669). Dashed lines represent the corresponding simulated spectra generated with the CEF model discussed in~\ref{Multiplets}, assuming no broadening.}
\label{Fig_RIXS-LD}
\end{figure}

In order to get an overview of the Cr electronic structure as seen by RIXS, we have acquired extended RIXS maps covering the Cr $L_{2,3}$ edges with both linear polarizations in grazing incidence geometry. The results are displayed in Fig.~\ref{Fig_RIXS-LD}\textbf{ab}. Additional high-statistics RIXS spectra, measured at 25~K with incident energies indicated by the white arrows in Fig.~\ref{Fig_RIXS-LD}\textbf{a}, are displayed in Fig.~\ref{Fig_RIXS-LD}\textbf{c} and compared against our crystal field model, which will be detailed in section~\ref{Multiplets} and in Appendix~\ref{sec:appCF}. We note that due to a scattering angle close to 90$^\circ$, the elastic peak intensity is strongly reduced for $\pi$ polarization. In both maps, two well defined vertical lines can be noticed, running parallel to the elastic peak, i.e. at a constant energy loss. These lines correspond to well-defined peaks (Fig.~\ref{Fig_RIXS-LD}\textbf{c} and Fig.~\ref{Fig_RIXS_LD1}) and vary in intensity as a function of the incident energy. Their resonant and Raman behavior as a function of excitation energy, as well as their position in energy, are characteristic of crystal electric field allowed transitions, corroborating the attribution of the two peaks to d-d excitations. Notably, these losses are stronger in the $\pi$-polarization map compared to the $\sigma$ one, matching the observed XLD behavior (Fig.~\ref{Fig_geo_and_LD}\textbf{a}).
Additionally, a broad and seemingly structureless intensity is observed at higher energy loss as diffused horizontal features, showing little to no dichroism.This feature exhibit some resonance with the incident energy and is present at both absorption edges, suggesting contributions from unresolved states, likely of charge transfer origin. Finally, the diagonal lines in this energy loss representation correspond to a fluorescent signal. The presence of a gap across the entire incident energy range supports the semiconducting nature of the material.

By looking at the spectra in Fig.~\ref{Fig_RIXS-LD}\textbf{c}, in most of them, the first of the two d-d excitations appears to be the strongest. The intensity ratio between the two main d-d excitations measured in $\pi$ polarization (green) shows substantial non-linear variations as a function of incident energy, with the first one resonating strongly at specific energies. This effect is much weaker in the $\sigma$ polarization (brown), but small variations are still observed, with the two excitations becoming comparable in intensity at certain incident energies. These observations, as well as some peaks asymmetry, suggest multiple contributions to both peaks, each resonating at different energies. The spectroscopic identification of these two peaks will be discussed in the next section, in view of our theoretical simulations. By comparing our model against the RIXS data, we note that it reproduces quite well the d-d positions, their linear dichroism and most intensity ratios. The main discrepancies can be observed in the stronger resonances observed in the first d-d for some excitation energies in $\pi$ configuration, and the observation of some dichroism in the broad features above 3~eV. Concerning this latter it should be pointed out that this spectral range show contributions from fluorescence emission, which cannot be captured by this theoretical model. By looking at the maps, it seems that some dichroism is indeed present in the fluorescent signal, but its interpretation is beyond the scope of this work.


\subsection{A Crystal Electric Field Model for CrSBr}\label{Multiplets}

\begin{figure}[tb]
\includegraphics[width=0.5\textwidth]{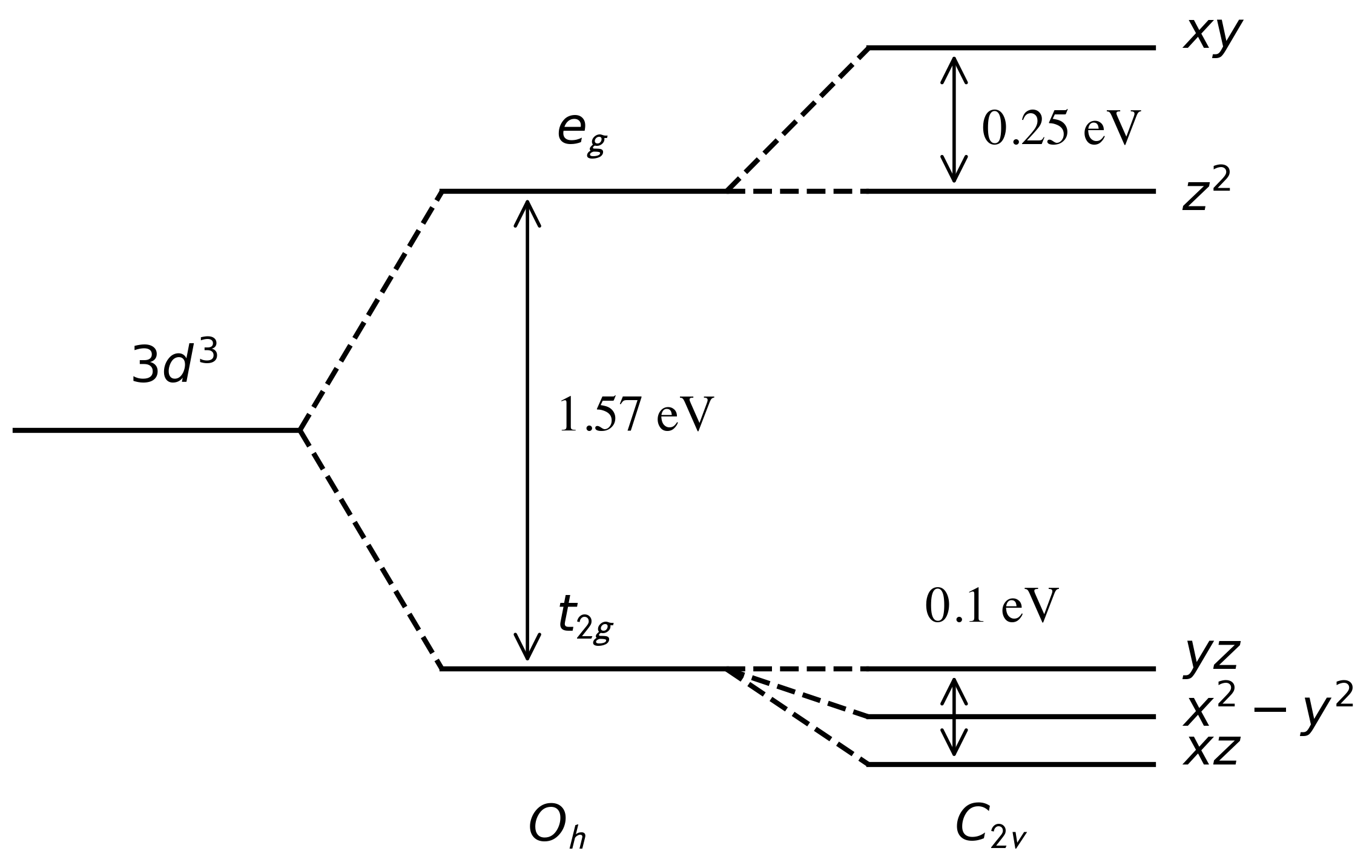}
\centering
\caption{Schematics of the 3\textit{d} orbitals splitting determined for Cr$^{3+}$ in CrSBr.}
\label{CEFscheme}
\end{figure}

To further understand the electronic structure of CrSBr, we conducted multiplet calculations using Quanty~\cite{Quanty1,Quanty2,Quanty3}, employing a Crystal Electric Field (CEF) model that captures the local environment of the Cr ions ($C_{2v}$). 
The detailed procedure for fixing the CEF parameters is described in Appendix ~\ref{sec:appCF}. Essentially, the position of the first d-d excitation sets a splitting of 1.57~eV between $t_\mathrm{2g}$ and $e_\mathrm{g}$ orbitals. No degeneracy being enforced by the point group, the hierarchy of the partially occupied $t_\mathrm{2g}$ and unoccupied $e_\mathrm{g}$ orbitals was established by comparing the polarization and angular dependence of both the XAS and RIXS measurements against simulations. The splitting of the $e_\mathrm{g}$ orbitals by 0.25~eV reproduces the shift of the $L_2$ observed in the XAS, as well as the broadening of the d-d excitations in the RIXS spectra. An optimal splitting of 0.1~eV was found, producing pre-peak features consistent with the observed XAS spectra. The best fit was obtained with the orbital order $E_{xz}$~\textless~$E_{xy}$~\textless~$E_{yz}$. Additionally, incorporating a DFT-predicted inversion of the $xy$ and $x^2-y^2$ orbitals (as reported in \cite{DFT_CrSBr_CEF} and Supplemental Material) and a magnetic exchange term of -0.013~eV~\cite{CrSBr_Scheie_magnons} was found to slightly improved the simulations. The complete list of parameters is provided in Table~\ref{Param_table} and Fig.~\ref{CEFscheme}.\\

\begin{table}[tb]
\begin{tabular}{|c|c|c|}
\hline
Parameters      & Initial State & Core-hole excited state \\ \hline
$F^2_{dd}$            & 5.2185 (48\%)    & 5.5659 (48\%)                  \\ \hline
$F^4_{dd}$            & 5.2920 (78\%)    & 5.6711 (78\%)                  \\ \hline
$F^2_{pd}$            & -                & 5.5464 (85\%)                  \\ \hline
$G^1_{pd}$            & -                & 3.3496 (70\%)                  \\ \hline
$G^3_{pd}$            & -                & 2.5851 (95\%)                  \\ \hline
$\zeta_{2p}$          & -                & 5.6678 (100\%)                 \\ \hline
$\zeta_{3d}$          & 0.0177 (50\%)    & 0.0234 (50\%)                  \\ \hline
$Ea2$ ($E_{xy}$)        & 1.82             & 1.82                           \\ \hline
$Ea1z2$ ($E_{z^2}$)      & 1.57             & 1.57                           \\ \hline
$Eb2$ ($E_{yz}$)        & 0.00& 0.00                           \\ \hline
$Ea1x2y2$ ($E_{x^2-y^2}$) & -0.05            & -0.05                          \\ \hline
$Eb1$ ($E_{xz}$)        & -0.10            & -0.10\\ \hline
$J_{ex}$             & -0.013            & -0.013                          \\ \hline
\end{tabular}
\caption{The parameters used in the CEF model are listed in the first column. The second column contains the values used to build the list of eigenstates in the initial and final states of the RIXS, while the last column contains values used for the intermediate state of the RIXS (final state of the XAS). All values are given in units of eV and percentages correspond to the scaling factors applied relative to the atomic Hartree-Fock values.}
\label{Param_table}
\end{table}

\begin{figure}
\includegraphics[width=0.48\textwidth]{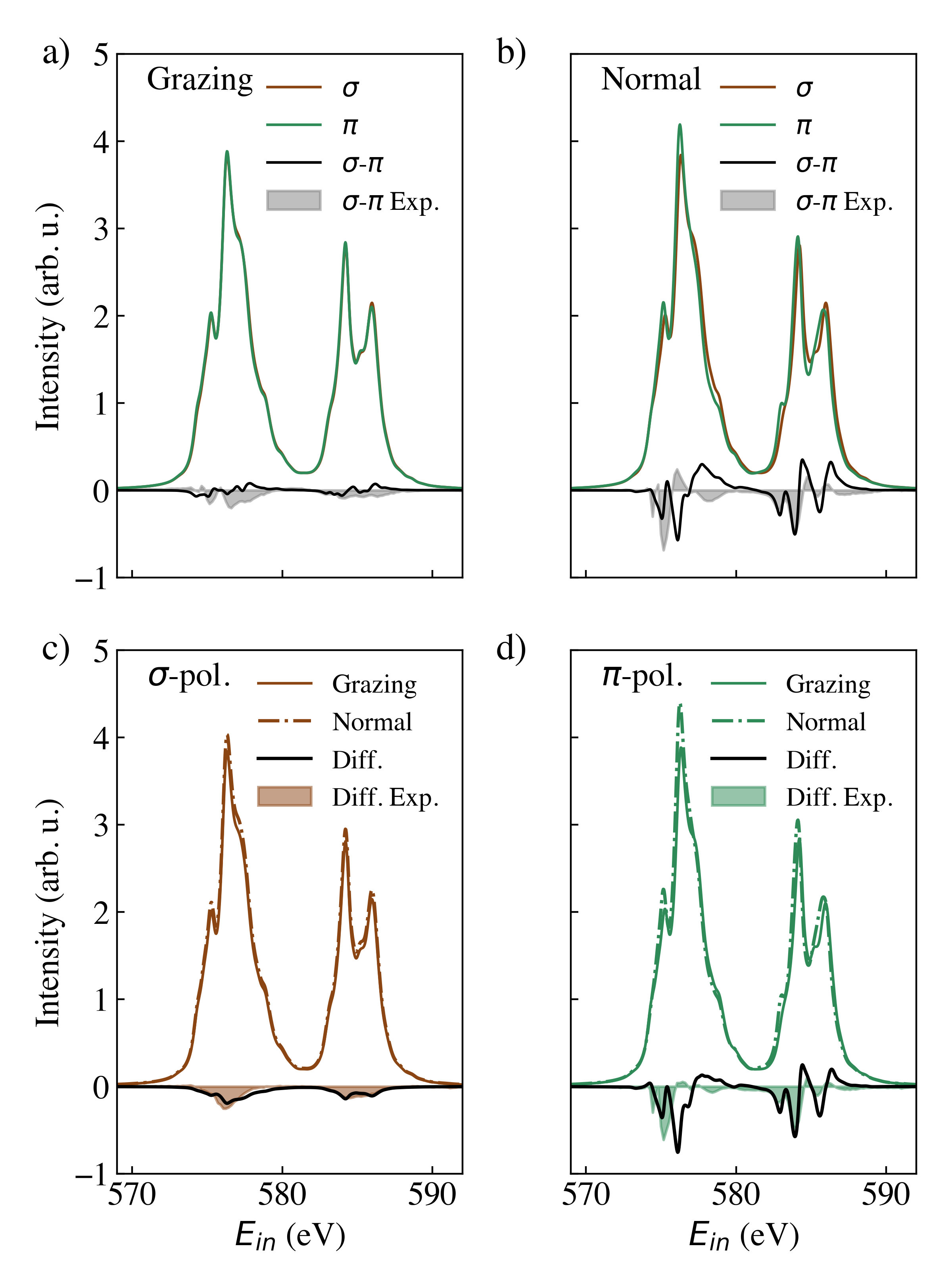}
\centering
\caption{Simulated XAS spectra and associated linear dichroism collected at RT for both grazing (a) and normal (b) incidence at the Cr $L_{2,3}$ edges. The $\pi$ ($\sigma$) polarization corresponds to an incident radiation's electric field oriented parallel (perpendicular) to the scattering plane. Comparison between simulated spectra in grazing and normal incidence for $\sigma$ (c) and $\pi$ (d) polarizations. The black lines and shaded areas correspond to the simulated and observed dichroism, respectively.}
\label{XAS_LD_sim}
\end{figure}

The calculated XAS spectra, shown in Fig.~\ref{XAS_LD_sim}, illustrate the simulated polarisation and angular dependence. As mentioned earlier, simultaneous reproduction of the linear dichroism at the $L_3$ and $L_2$ edges proved challenging despite exploring a wide parameter range. However, some key features are reasonably captured. The linear dichroism amplitude increases from grazing to normal incidence (Fig.~\ref{XAS_LD_sim}\textbf{ab}), with the slight dominance of $\pi$ polarization in grazing geometry also observed, albeit less pronounced than in the experiment. In normal geometry, the $L_2$ behavior is reasonably well captured, particularly in the lower half where a shift was noted. However, the $L_3$ dichroism is poorly accounted for, likely due to strong hybridization between the Cr orbitals and its ligands, which our model does not include. Tests using higher symmetry ($D_{4h}$) suggest that adding a charge transfer term could improve this aspect. However, incorporating such a term within the lower $C_{2v}$ symmetry, combined with two ligands, would significantly increase the number of parameters and is left for future studies. The angular dependence is more successfully reproduced, especially for the $\sigma$ polarization. The $\pi$ polarization remains poorly modeled around the $L_3$ edge but is again remarkably accurate on the left side of the $L_2$. Additionally, upscaling the normal geometry simulation by 5~\% aligns better with the observed dichroic amplitude, likely accounting for the geometrical dependence of the incident beam’s footprint on the sample.\\

\begin{figure}
\includegraphics[width=0.48\textwidth]{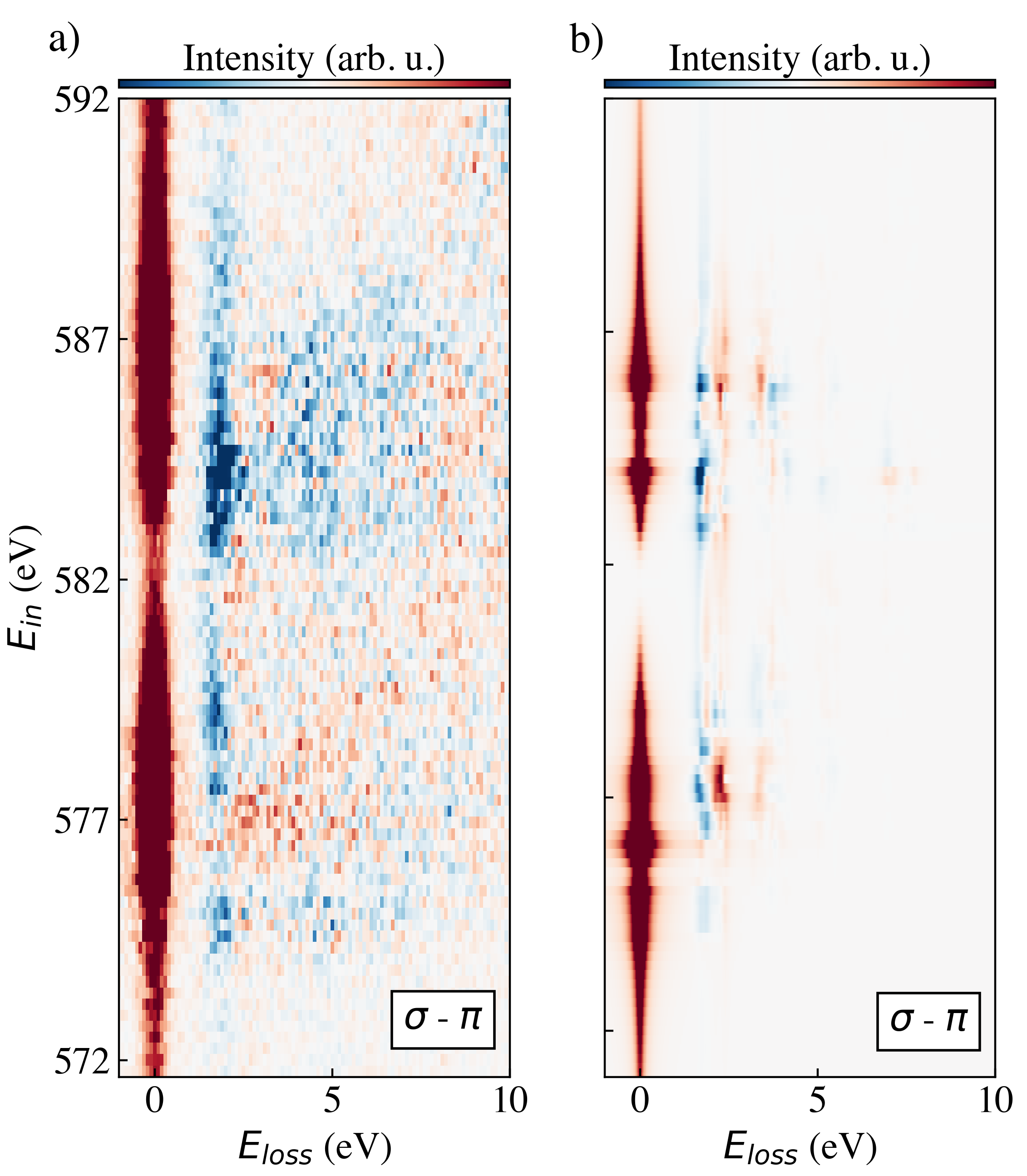}
\centering
\caption{Comparison between experimental (a) and simulated (b) RIXS linear dichroism map in grazing geometry at RT. The dichroism maps are generated by subtracting maps obtained with the two linear polarizations ($\sigma - \pi$). The red denotes regions dominated by the $\pi$ polarization, whereas the blue ones are dominated by the $\sigma$ polarisation.}
\label{RIXS_LD_sim}
\end{figure}

The comparison between the experimental and simulated (Fig.~\ref{RIXS_LD_sim}\textbf{ab}) RIXS dichroism maps shows the capacity of our model to reproduce the observed Raman losses, particularly the dichroism of the first d-d excitation, with a clear dominance of the $\pi$ polarization across most of the incident energy range. We also observe some positive dichroism regions at some incident energies, mainly for the second d-d excitation, which aligns with experimental observations. The angular dependence of the dichroism around the $L_{3}$ pre-peak (Fig.\ref{Fig_RIXS_LD1}) is also reproduced, with a notable increase of the first d-d excitation's dichroism in normal incidence (see Supplemental Material). The agreement is also satisfactory when looking at the RIXS spectra from Fig.~\ref{Fig_RIXS-LD}\textbf{c}, with the simulation capturing the position, intensities and dichroism of the main d-d excitations, as well as their apparent doubling. Trying to relate excitations observed in the RIXS and the photoluminescence reported in CrSBr, our model suggests that the second contribution to the first peak, perhaps giving rise to the second bright exciton, corresponds to a $^4A_\mathrm{2g}$ $\rightarrow$ $^4T_\mathrm{2g}$ electronic transition partially filling the uppermost ($xy$) orbital.

\subsection{X-ray Excited Optical Luminescence}\label{XEOL_part}

\begin{figure}[tb]
\includegraphics[width=0.4\textwidth]{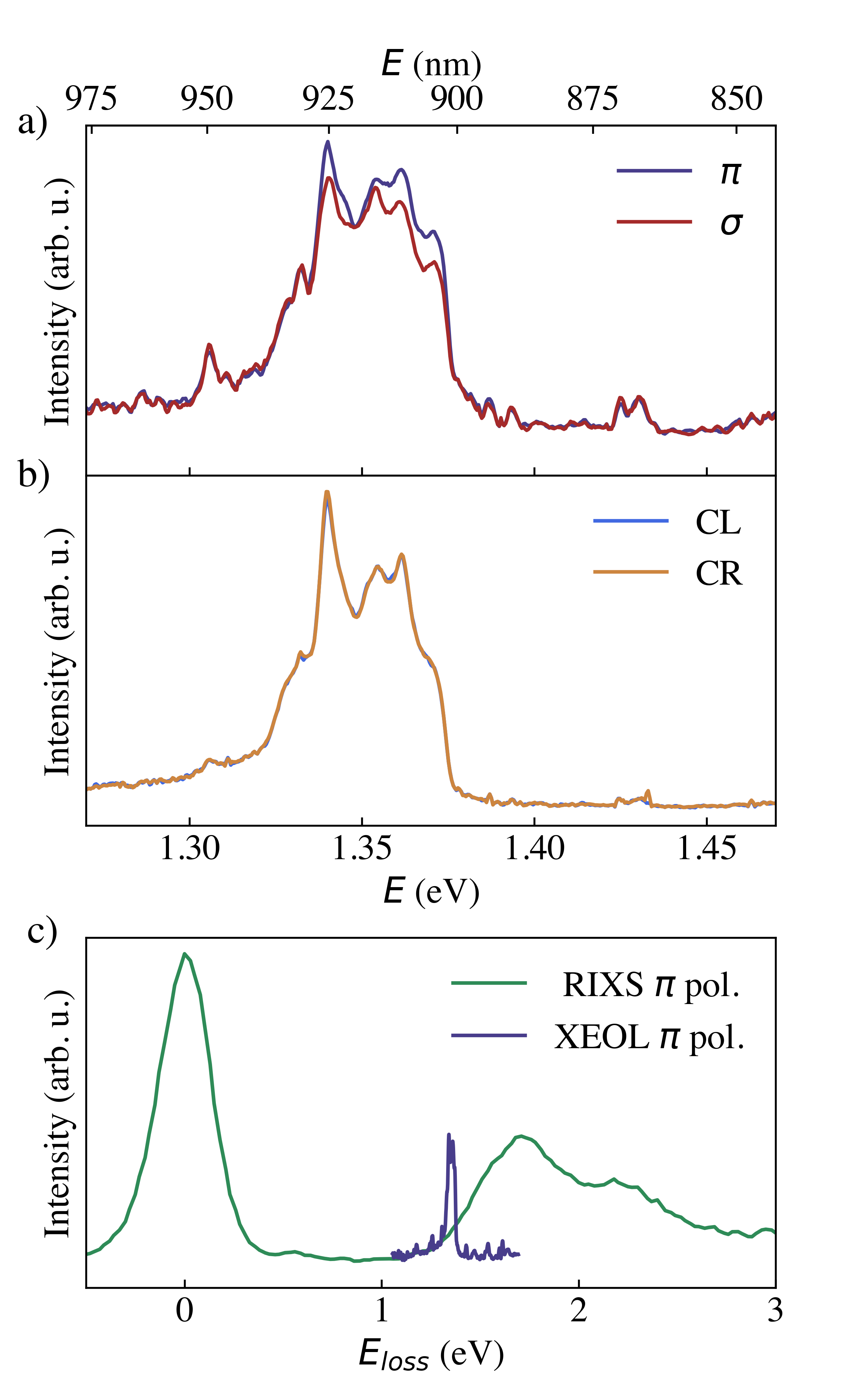}
\centering
\caption{XEOL spectra measured at 20~K using both linear (a) and circular (b) polarizations at an incident radiation energy of 579.6~eV. (c) Comparison between XEOL spectrum from panel (a) and RIXS one from Fig.~\ref{Fig_RIXS_LD1} acquired using the $\pi$ polarization in grazing incidence.}
\label{Fig_PL}
\end{figure}

In this section we discuss the photoluminescence (PL) signal generated by the sample at 20~K and under x-ray radiation near the $L_{3}$ edge. Using an incident radiation of 579.6~eV, a structured peak is observed, centered at approximately 1.35~eV, as pictured in Fig.~\ref{Fig_PL}\textbf{a}. This emission is consistent with the photoluminescence of excitonic origin observed in this compound using other techniques~\cite{CrSBr_NatMat,CrSBr_NatNano,MultiPL}. The spectral shape of the PL changes slightly between $\sigma$ and $\pi$ polarizations and is weaker for the former one. This reduction in PL observed with $\sigma$ polarization (electric field along the \textbf{a}-axis) is likely due to the same mechanism that causes the absence of a PL signal in optical spectroscopy experiments with the same polarization orientation. This absence is attributed to a dipole-forbidden matrix element for this configuration, whereas it is allowed for a polarization along the \textbf{b}-axis. In our case, this condition is relaxed for the absorption step as we excite from a 2$p$ orbital.


The broad shape of the signal is likely due to the excitation of exciton-polaritons as already observed in bulk samples using other optical spectroscopy techniques~\cite{MultiPL,SecondPL}. It is important to note that the method used to acquire these spectra differs from what has been used on CrSBr to measure its photoluminescence signal up to now, which was relying on irradiation using IR/visible light and therefore on direct transitions between the valence and conduction bands.\\
The same spectra measured using both circular polarizations are displayed Fig.~\ref{Fig_PL}\textbf{b}, showing no dichroism. The structure of the peak is slightly modified, suggesting a different cross section between linearly and circularly polarized incident radiations.\\
Comparing the RIXS and XEOL spectra, as illustrated in Fig.~\ref{Fig_PL}\textbf{c}, we note that the lowest excitation in the RIXS spectra is 0.22~eV above the lowest bright exciton seen in the XEOL and previous reports~\cite{NanoLetter_Ziebel,CrSBr_NatNano}. The observed photoluminescence is therefore unlikely to result from chromium intra-atomic electronic transitions, in line with current interpretations~\cite{CrSBr_NatMat,CrSBr_NatNano,Wang_CrSBr}. However, we cannot exclude that the secondary photoluminescence signal, reported at 1.75~eV~\cite{CrSBr_NatMat,SecondPL}, has a potential d-d transition origin, as it matches the second contribution to the first d-d transition we observe in the RIXS spectra. We also note a difference in width between the photoluminescence and the RIXS, with the former one significantly narrower than the d-d excitations due to the very different experimental resolution of the two experiments. This remark might also explain the absence of any structure related to the lowest exciton in the RIXS spectra, considering our experimental conditions.




\section{Conclusion}
In this study, we provide new insights into the nature of Cr$^{3+}$ electronic states in CrSBr and their interaction with the distorted local environment of Cr. To achieve this, we combine X-ray absorption spectroscopy, resonant inelastic X-ray scattering, and X-ray excited optical luminescence measurements, complemented by multiplet calculations. 
While XEOL confirms the presence of excitonic emission exhibithing weak polarization dependence, XAS and RIXS measurements reveal significant linear dichroism consistent with a distorted octahedral environment around the Cr ions. RIXS data, which align with the semiconducting nature of CrSBr, show d-d excitations typically observed in Cr$^{3+}$ systems, with the two major peaks corresponding to transitions from the ground state quartet to the first two excited quartet states ($^4A_\mathrm{2g}$ $\rightarrow$ $^4T_\mathrm{2g}$ and $^4A_\mathrm{2g}$ $\rightarrow$ $^4T_\mathrm{1g}$). 
By correlating RIXS and XEOL measurements, we identify the lowest discernible d-d excitation in RIXS approximately 0.2~eV higher than the energy of the lowest observed bright exciton, making it unlikely to correspond to the same electronic transition. 
Our multiplet model, which incorporates the actual $C_{2v}$ symmetry, successfully reproduces many aspects of the experimental XAS and RIXS spectra, including angular dependence and the main dichroic trends. However, the complex behavior of the XLD at the $L_3$ edge suggests strong hybridization effects between Cr 3\textit{d} orbitals and their surrounding ligands, which are not captured by our model. Accordingly, RIXS maps provide evidence of non negligible charge transfer excitations in CrSBr. Future studies should explore more advanced models, potentially including charge transfer effects, to better account for the strong orbital hybridization present here.\\ 
Overall, the comprehensive analysis presented in this work clarifies the electronic structure of CrSBr, revealing the hierarchy of 3\textit{d} states near the Fermi level and their relationship with luminescence. These insights provide a valuable foundation for future studies exploring CrSBr as an active material in spintronic and optoelectronic devices, where precise control over electronic and optical properties is essential for advanced functionalities.\\

\appendix
\label{appendix}
\section{Crystal Electric Field Model: Procedure Used to Fix the Parameters}
\label{sec:appCF}

\begin{figure}[b]
\includegraphics[width=0.48\textwidth]{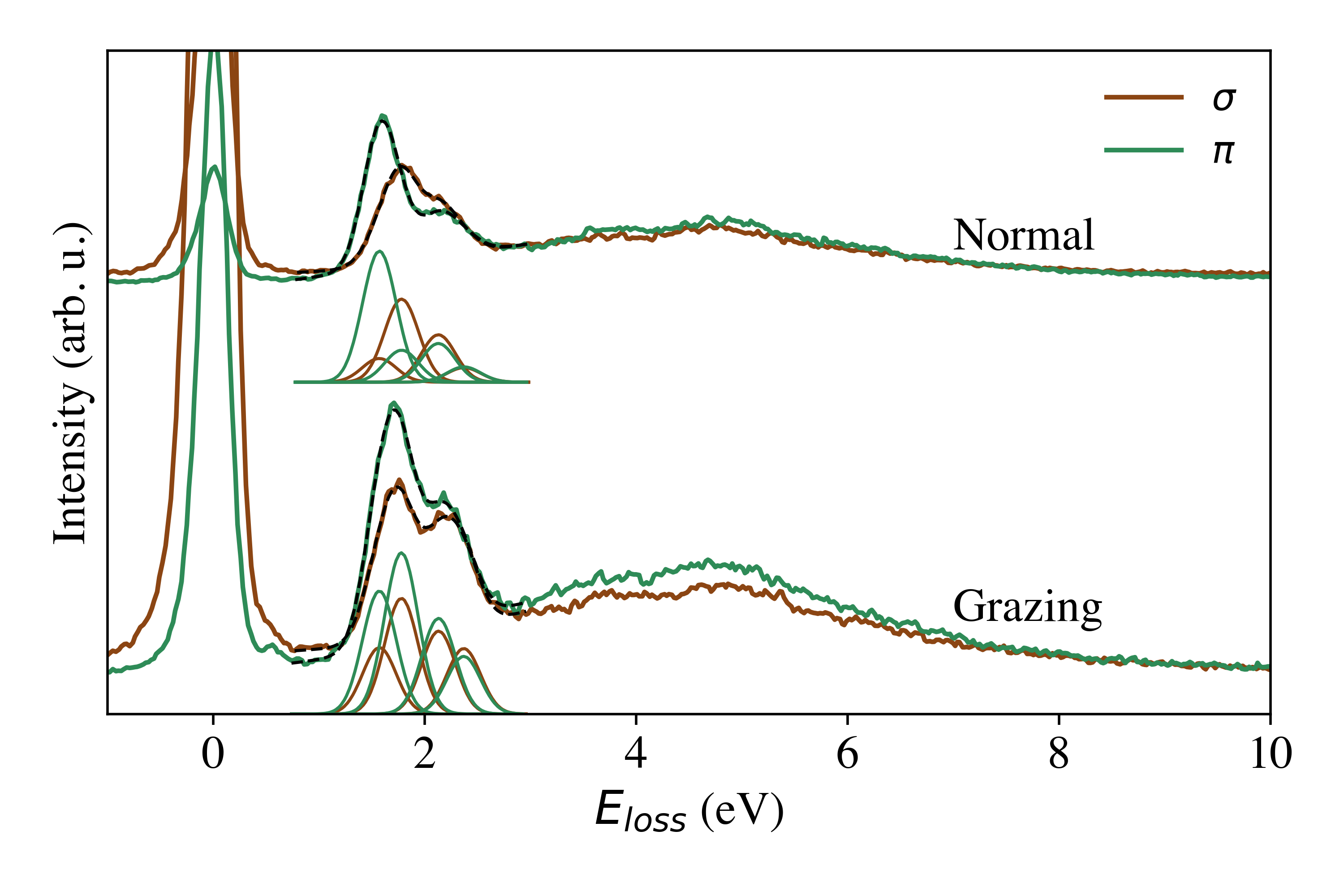}
\centering
\caption{RIXS spectra collected at RT with both linear polarizations for grazing and normal incidence. The incident energy was chosen according to the orange arrows in Fig.~\ref{Fig_geo_and_LD}. The four Individual Gaussian contributions used to fit the excitations are displayed below each spectrum, using matching colors, while the global fits are shown as dashed black lines.}
\label{Fig_RIXS_LD1}
\end{figure}

In order to allow us to accurately determine and refine the parameters of our Hamiltonian, high-statistics RIXS spectra were acquired in normal and grazing incident geometries with both $\pi$ and $\sigma$ linear polarizations. The result is presented in Fig.~\ref{Fig_RIXS_LD1}. These spectra were obtained at the same incident energy of 575.2~eV, corresponding to the maximum of the dichroism observed in XAS using the normal geometry, as indicated by the orange arrows in Fig.~\ref{Fig_geo_and_LD}(\textbf{ab}). 
The normal incidence measurements highlight a dichroic behavior of the d-d excitations, which is weaker in the grazing geometry. Four Gaussian functions, with their width fixed at the experimental resolution, were required to reproduce these excitations (two per observed peak, as depicted in Fig.~\ref{Fig_RIXS_LD1}). This apparent doubling of the crystal field levels can be explained by the distortion of the crystal electric field, away from the octahedral symmetry, in particular due to the degeneracy lifting of the Cr$^{3+}$ $e_\mathrm{g}$ orbitals. A first fit of the four spectra was performed without constraints on the Gaussian's positions. Subsequently, their positions were fixed to the average values obtained across the four spectra, with an error defined as their variance. This procedure places the lowest excitation at 1.57$\pm$0.06~eV, followed by excitations at 1.78$\pm$0.06~eV, 2.13$\pm$0.04~eV and 2.37$\pm$0.04~eV.

\begin{figure}[tb]
\includegraphics[width=0.5\textwidth]{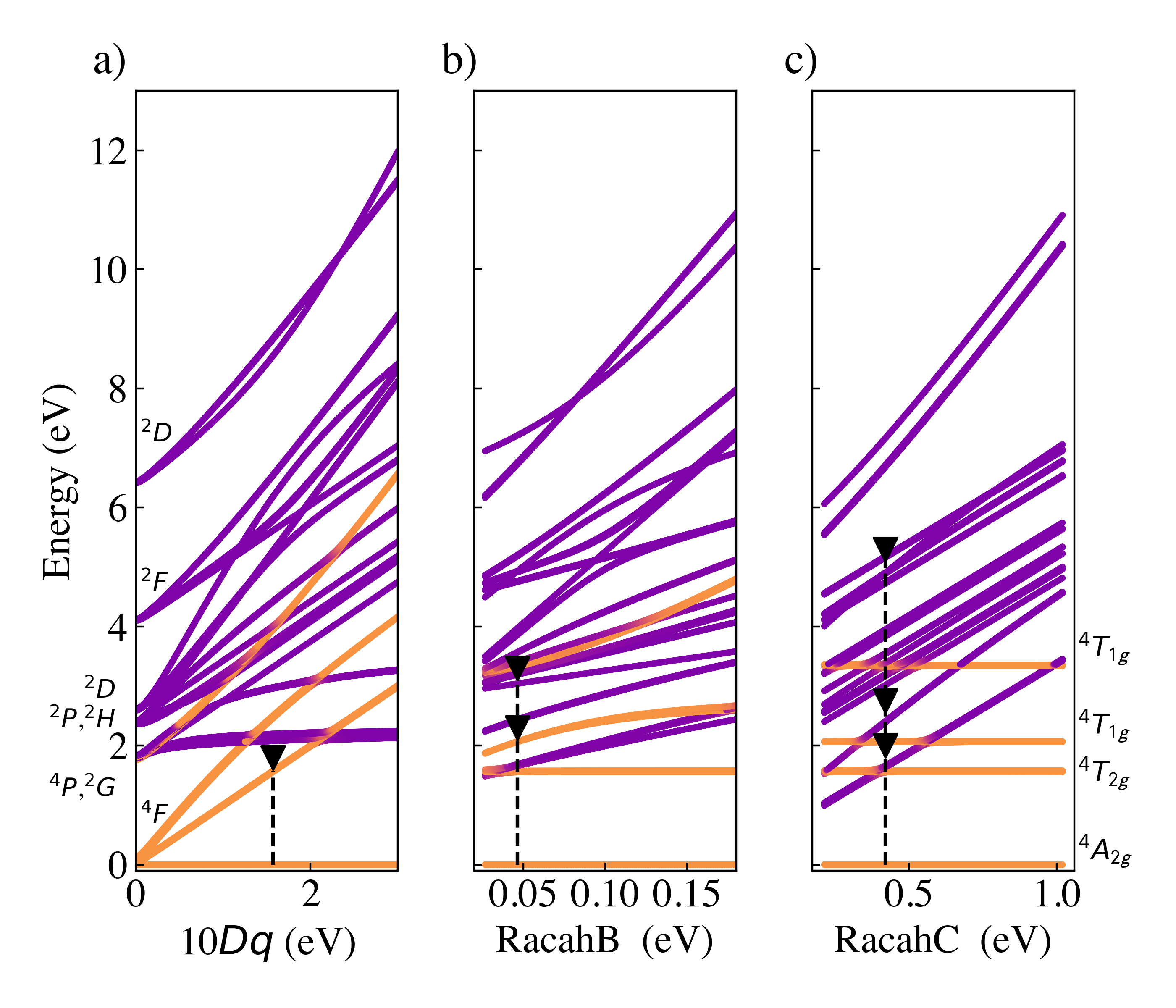}
\centering
\caption{Tanabe-Sugano diagrams for Cr$^{3+}$, varying 10$Dq$ (a), RacahB (b) and RacahC (c). Orange and violet lines depict quartet and doublet states, respectively. The states used to fix the parameters are indicated by the black dashed lines and triangles. For clarity, only the quartet states and multiplet state before crystal field splitting are indicated, while a more complete list of states can readily be found in the literature~\cite{Hunault_Ruby,Shao_Cr_RIXS}.}
\label{T-S_diag}
\end{figure}

We then compare the RIXS excitations to Tanabe-Sugano diagrams for Cr$^{3+}$ in an octahedral ($O_h$) symmetry, as drawn in Fig.~\ref{T-S_diag}. The Hamiltonian used incorporates intra-atomic Coulomb interactions between 3\textit{d} electrons ($F^2_{dd}$ and $F^4_{dd}$), CEF splitting (10$Dq$), which separates the 3\textit{d} orbitals into $t_\mathrm{2g}$ and $e_\mathrm{g}$ levels, and the Spin-Orbit coupling (SOC).
We initialize the Hamiltonian with Coulomb interactions parameters (Slater integrals) set at 80~\% of their Hartree-Fock values and the SOC scaled to 50~\%, following standard practice in this kind of system~\cite{Hunault_Ruby,Shao_Cr_RIXS,CrI3_PL_Peng}. The system's eigenstates and associated energy levels are then computed as a function of 10$Dq$, as shown in Fig.~\ref{T-S_diag}\textbf{a}. To align with experimental results, we consider only positive 10$Dq$ values, maintaining a $^4A_\mathrm{2g}$ ground state with S~=~3/2. The focus here is on the energy of the first excited quartet state (S~=~3/2), which determines the values of 10$Dq$ and is expected significantly contribute to one of the two observed peaks.

\noindent
Both scenarios were tested, revealing that assigning the first quartet state ($^4T_\mathrm{2g}$) to the second peak led to RIXS spectra where the second peak became dominant across most incident energies, contradicting our observations. Consequently, we align the first quartet state with the first excitation, consistent with approaches used in chromium trihalides~\cite{Shao_Cr_RIXS}, setting 10$Dq$ at 1.57~eV (Fig.~\ref{T-S_diag}\textbf{a}). We then adjust the Slater integrals via the RacahB and RacahC parameters (where RacahB~=~(9$F^2_{dd}$-5$F^4_{dd}$)/441 and RacahC~=~5$F^4_{dd}$/63) to match other eigenstates with the observed peaks (Fig.~\ref{T-S_diag}\textbf{bc}). Based on a reasonable range of values and on computed intensities, a satisfactory match was obtained with RacahB~=~0.0465~eV and RacahC~=~0.42~eV. Under these conditions, the first peak is attributed to both a quartet ($^4T_\mathrm{2g}$) closely followed by two doublets ($^2E_\mathrm{g}$ and $^2T_\mathrm{1g}$), while the second peak corresponds to the next quartet ($^4T_\mathrm{1g}$) followed by a doublet ($^2T_\mathrm{2g}$). This configuration also positions the last quartet and several doublets around 3.2~eV, aligning with some weak peaks arising from the broad fluorescent and charge transfer emissions  (Fig.~\ref{T-S_diag}\textbf{c}). Additionally, a cluster of doublet states at approximately 4.7~eV conveniently aligns with other structures observed in several spectra.\\

\begin{figure}[tb]
\includegraphics[width=0.48\textwidth]{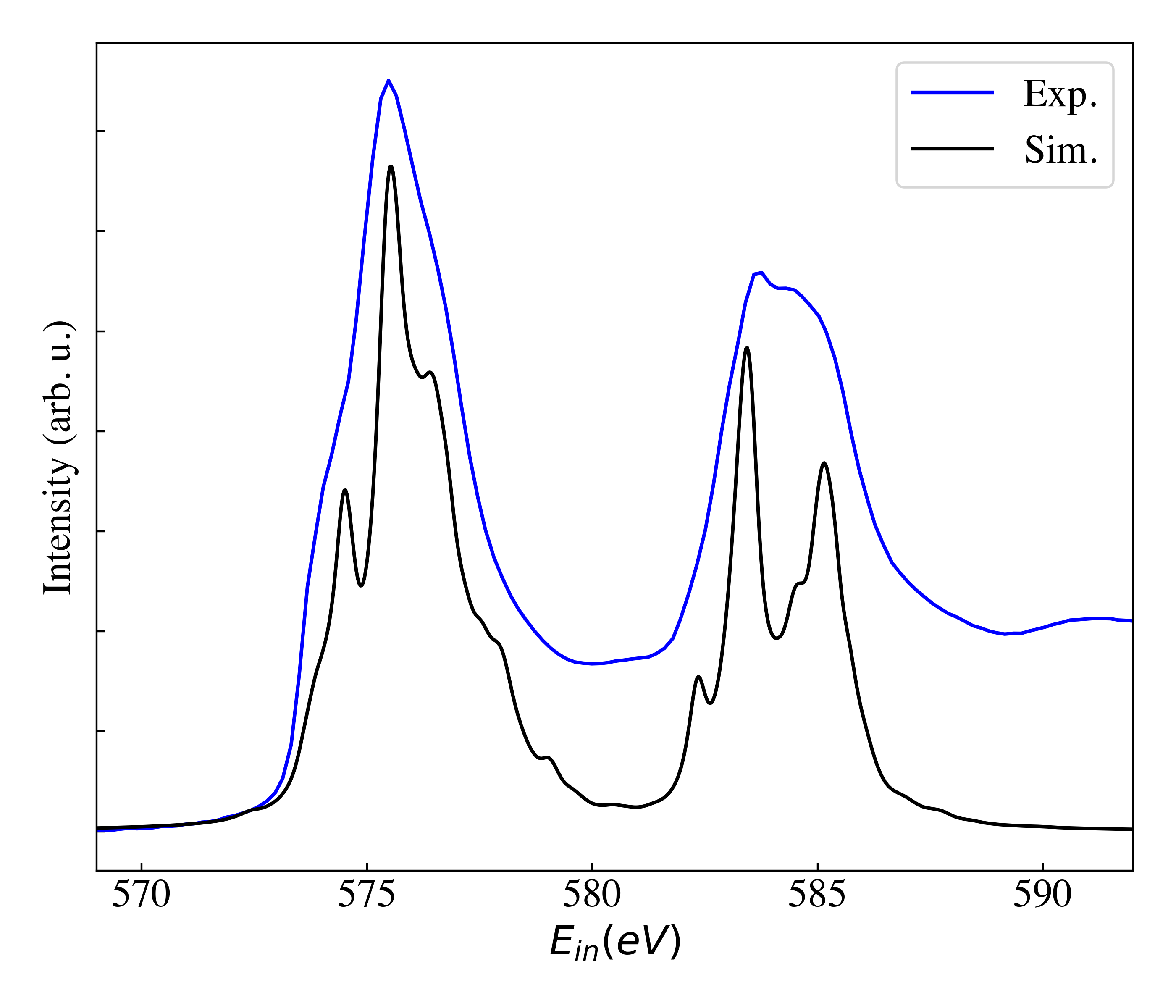}
\centering
\caption{Comparison between the measured (blue) and simulated (black) isotropic XAS spectra. The experimental isotropic spectrum was obtained by averaging the XAS collected in grazing incidence with both circular polarizations at RT.}
\label{XAS_iso}
\end{figure}

Next, we determine the Hamiltonian's parameters for the final (intermediate) state of the XAS (RIXS). As is customary in these systems, the scalings of the $F^2_{dd}$ and $F^4_{dd}$ Slater integrals and SOC remain consistent with the initial state, while their Hartree-Fock values account the extra \textit{d} electron. This leaves three parameters: $F^2_{pd}$, $G^1_{pd}$ and $G^3_{pd}$, representing the Coulomb interaction between the core-hole and the valence electrons. Optimal agreement with the observed XAS peak hierarchy is achieved by scaling these parameters to 85~\%, 70~\% and 95~\%, respectively (see Fig.~\ref{XAS_iso}).

However, the observed linear dichroism cannot be reproduced in the $O_h$ symmetry approximation. The presence of linear dichroism proves indeed a deviation from this simple symmetry. In order to go further with our modeling, we therefore leverage Quanty's flexibility to implement a CEF model reflecting the actual point group symmetry experienced by the Cr ions, namely $C_{2v}$. In this symmetry, no orbital degeneracy is enforced, allowing distinct energy levels for each 3\textit{d} orbitals. Previous theoretical and experimental studies~\cite{Wang_CrSBr,DFT_CrSBr_CEF,CrSBr_NatComm_Klein,Bianchi_ARPES,CrSBr_Linhart} suggest a small deviation from $O_h$ symmetry, leading to a split of both $t_\mathrm{2g}$ and $e_\mathrm{g}$ orbitals, while keeping the three lowest well-separated from the upper two. Thus, we retain 10$Dq$~=~1.57~eV as gap between the highest partially occupied orbital and the lowest unoccupied one.\\
To set the splitting of the $e_\mathrm{g}$ orbitals ($z^2$ and $x^2-y^2$), we introduce a 0.25~eV splitting, as suggested by the observed $L_2$ edge shift (see \ref{XAS_section}). This choice is also coherent with the doubling of the d-d excitations observed in the RIXS spectra (Fig.~\ref{Fig_RIXS_LD1}). To determine the hierarchy of the two former $e_\mathrm{g}$ orbitals, we evaluate both possible configurations. Simulations of the XAS spectra, however, remain inconclusive, as neither configuration consistently captures the dichroism across both edges. Nevertheless, the $L_2$ dichroism, especially the observed shift, is better reproduced with $z^2$ as the lowest of the two $e_\mathrm{g}$ orbitals. Further simulations of the RIXS maps confirm this arrangement as it accurately models the dichroism, particularly for the first d-d excitation, which is not achieved with the alternative configuration. The ordering of the remaining three orbitals was found to have minimal impact on these results. This $e_\mathrm{g}$ orbitals configuration aligns with DFT calculations (Supplemental Material and~\cite{Wang_CrSBr,DFT_CrSBr_CEF}), which also predict an inversion of these two orbitals between spin-up and spin-down states, a detail that our CEF model may not fully capture, potentially explaining discrepancies in the XAS linear dichroism, which we will discuss later.\\
Finally, we determine the splitting of the $t_\mathrm{2g}$ orbitals. While our DFT calculations suggest a separation of around 0.3~eV between the highest and lowest half-filled orbitals, XAS simulations indicate that splittings beyond 0.15~eV introduce significant pre-peak features inconsistent with the experimental data. Nevertheless, a splitting of 0.1~eV led to pre-peak features consistent with the observed XAS spectra. The best agreement between our model and all of our experimental data was eventually obtained with the $t_\mathrm{2g}$ orbital order $E_{xz}$~\textless~$E_{xy}$~\textless~$E_{yz}$.

\begin{acknowledgements}
This work was supported by a public grant overseen by the French National Research Agency (ANR) as part of the TRIXS project ANR-19-CE30-0011, and by the BONASPES project (ANR-19-CE30-0007). Z.S. and J.R. were supported by ERC-CZ program (project LL2101) from Ministry of Education Youth and Sports (MEYS) and by the project Advanced Functional Nanorobots (reg. No.~CZ.02.1.01/0.0/0.0/15\_003/0000444 financed by the EFRR). Finally, this work is part of the IMPRESS project that has received funding from the HORIZON EUROPE framework program for research and innovation under grant agreement n. 101094299.
\end{acknowledgements}

\bibliography{main.bbl}

\end{document}